\documentclass{elsarticle}
\usepackage{comment}

\usepackage{amssymb}
\usepackage{amsmath}

\usepackage{graphicx}
\usepackage{color}
\usepackage{ulem}
\usepackage{xcolor}
\DeclareRobustCommand{\erase}{\bgroup\markoverwith{\textcolor{red}{\rule[.5ex]{2pt}{0.4pt}}}\ULon}

\usepackage{doi}
\usepackage{hyperref}

\usepackage{subcaption}
\usepackage{multirow}

\journal{Nuclear Instruments and Methods in Physics Research Section A}

\begin{document}
\begin{frontmatter}



\title{Simulation Tool Development and Sensitivity Analysis of $^{160}$Gd Double Beta Decay Search by the PIKACHU Project} 


\author[address1]{T.~Omori\corref{mycorrespondingauthor}} 
\author[address2]{T.~Iida\corref{mycorrespondingauthor}}
\cortext[mycorrespondingauthor]{Corresponding author}
\ead{omori@hep.px.tsukuba.ac.jp, tiida@hep.px.tsukuba.ac.jp}

\author[address2,address3]{N. Hinohara}
\author[address1]{K. Takahashi}
\author[address4]{K.~Fushimi}
\author[address5,address6]{A.~Gando}
\author[address6]{K.~Hosokawa}
\author[address1]{S.~Ishidate}
\author[address1]{M.~Ishigami}
\author[address7,address8]{K.~Kamada}
\author[address6]{K.~Mizukoshi}
\author[address8]{Y.~Shoji}
\author[address1]{H.~Suzuki}
\author[address7,address8]{M.~Yoshino}

\address[address1]{Graduate School of Science and Technology, University of Tsukuba,  Tsukuba, Ibaraki, 305-8571, Japan}
\address[address2]{Institute of Pure and Applied Sciences, University of Tsukuba,  Tsukuba, Ibaraki, 305-8571, Japan}
\address[address3]{Center for Computational Sciences, University of Tsukuba, Tsukuba, Ibaraki, 305-8577, Japan}
\address[address4]{Division of Science and Technology, Tokushima University, 2-1 Minami Josanjima-cho Tokushima city, 
Tokushima, 770-8506, Japan}
\address[address5]{Department of Human Science, Obihiro University of Agriculture and Veterinary Medicine, Obihiro, Hokkaido, 080-8555, Japan}
\address[address6]{Research Center for Neutrino Science, Tohoku University, Sendai, 980-8578, Japan}
\address[address7]{New Industry Creation Hatchery Center, Tohoku University, Sendai, Miyagi 980-8579, Japan}
\address[address8]{C\&A Corporation, 1-16-23 Ichibancho, Aoba-ku, Sendai, Miyagi, 980-0811, Japan}


\begin{abstract}
Neutrinoless double beta decay (0$\nu$2$\beta$) has been investigated as a physical process that can provide evidence for the Majorana nature of neutrinos. The theoretical predictions of the 0$\nu$2$\beta$ rate are subject to significant uncertainty, primarily due to nuclear matrix elements (NME). To reduce this uncertainty, experimental measurements of the half-lives of two-neutrino double beta decay (2$\nu$2$\beta$) in various nuclei are essential as a benchmark for NME calculations. The PIKACHU (Pure Inorganic scintillator experiment in KAmioka for CHallenging Underground sciences) project searches for the previously unobserved 2$\nu$2$\beta$ of $^{160}$Gd, employing Ce-doped Gd$_{3}$Ga$_{3}$Al$_{2}$O$_{12}$ (GAGG) single crystals. In the Phase~1 experiment, we aim to improve the current lower limit on the 2$\nu$2$\beta$ half-life of $^{160}$Gd, set at 1.9 $\times$ 10$^{19}$ years (90$\%$ C.L.) by a prior study using a Ce-doped Gd$_2$SiO$_5$ (GSO) crystal. Ultimately, in Phase~2, the project seeks to achieve a sensitivity surpassing the theoretical prediction of 7.4 $\times$ 10$^{20}$ years, enabling the potential discovery of the $^{160}$Gd 2$\nu$2$\beta$. Understanding and evaluating backgrounds are crucial to determining the experimental sensitivity. In this paper, we describe the development of background models based on GEANT4 simulations. The modeled backgrounds are contributions from uranium and thorium decay chains, $^{40}$K present in GAGG, $^{40}$K and $^{208}$Tl $\gamma$-rays from outside of GAGG. Additionally, we developed models for both 2$\nu$2$\beta$ and 0$\nu$2$\beta$ by implementing the theoretical kinematics of two-electron emission in double beta decay in the GEANT4 simulation. As a result, our background models successfully reproduced the measured background spectrum through fitting. By generating pseudo background spectra expected in Phase~1 and analyzing them with the combined background and 2$\nu$2$\beta$ models, we evaluated the 2$\nu$2$\beta$ sensitivity of Phase~1 to be 2.64~$\times$~10$^{19}$ years (90$\%$ C.L.). This paper presents the development of these simulation models and the expected sensitivities for both Phase~1 and Phase~2 based on the pseudo data analyses. 
\end{abstract}



\begin{keyword}
Majorana neutrino \sep Double beta decay \sep Nuclear matrix element \sep Gadolinium 160 \sep GAGG scintillator \sep GEANT4


\end{keyword}

\end{frontmatter}




\section{Introduction}
\label{Introduction}
The neutrinoless double beta decay (0$\nu$2$\beta$) is a physical process beyond the framework of the Standard Model that violates the lepton number conservation law. The 0$\nu$2$\beta$ provides conclusive evidence for the Majorana nature of neutrinos, wherein neutrinos are identical to their own antiparticles. The half-life of 0$\nu$2$\beta$ ($T_{1/2}^{0\nu}$) is inversely proportional to the square of the effective electron neutrino mass ($m_{\beta\beta}$), offering crucial insights into the absolute neutrino mass. In contrast, the two-neutrino double beta decay (2$\nu$2$\beta$), a process allowed within the Standard Model, has been experimentally observed in several nuclei \cite{2nbb}. Many research groups are actively measuring the half-lives of 2$\nu$2$\beta$ ($T_{1/2}^{2\nu}$) in various double beta decay (2$\beta$) nuclei, as these measurements play a key role in improving the precision of nuclear matrix elements (NME) calculations \cite{NME}. NME, a nuclide-specific theoretical parameter that connects $T_{1/2}^{0\nu}$ and $m_{\beta\beta}$, carries significant uncertainties depending on the calculation method employed. Reducing these uncertainties is essential for enhancing the reliability of theoretical predictions and experimental interpretations in 0$\nu$2$\beta$ studies.\\
The PIKACHU (Pure Inorganic scintillator experiment in KAmioka for CHallenging Underground sciences) project focuses on investigating 2$\beta$ of $^{160}$Gd. $^{160}$Gd is a 2$\beta$ candidate with a high natural abundance (21.9$\%$) and a low $Q$-value (1.730~MeV \cite{Qval}) compared to other candidates. These characteristics suggest that an abundance of $^{160}$Gd can be utilized in experiments, highlighting the critical need to suppress background radiation with  energy near the $^{160}$Gd $Q$-value, arising from environmental radioactivity. The most sensitive 2$\beta$ search for $^{160}$Gd to date was conducted in Ukraine using a 2-inch square Ce-doped Gd$_{2}$SiO$_{5}$ (GSO) crystal containing 103.6 g of $^{160}$Gd \cite{Ukraine}. The sensitivity of this search was limited by background noise from uranium and thorium (U/Th) decay chains present in the GSO crystal. Consequently, while the 2$\nu$2$\beta$ was not observed, the lower limits at 90$\%$ C.L. for the $T_{1/2}^{2\nu}$ and $T_{1/2}^{0\nu}$ were set at 1.9 $\times$ 10$^{19}$ years and 1.3 $\times$ 10$^{21}$ years, respectively. \\
The PIKACHU project utilizes Ce-doped Gd$_{3}$Ga$_{3}$Al$_{2}$O$_{12}$ (GAGG) scintillator crystals for the 2$\beta$ search. GAGG is an inorganic scintillator for which large single crystal growth up by Czochralski method has been well-established \cite{Kochurikhin}. The GAGG crystal with dimensions of 65 mm in diameter and 145 mm in length contains approximately 355 g of $^{160}$Gd. Furthermore, GAGG has a superior light yield ($\sim$ 50,000 photons/MeV) compared to GSO ($\sim$ 10,000 photons/MeV), and its pulse shape discrimination (PSD) capability for particle identification \cite{Tamagawa} allows for effective removal of $\alpha$-ray background. However, the development of low-radioactivity GAGG remains an ongoing and critical research challenge.\\
The theoretical prediction of the 2$\beta$ half-life is associated with substantial uncertainty primarily due to NME. There are two typical predictions for the $T_{1/2}^{2\nu}$ in $^{160}$Gd: 9.6 $\times$ 10$^{21}$ years derived from the NME calculated in Ref. \cite{Hirsch}, and 7.4 $\times$ 10$^{20}$ years derived from the NME calculated using the quasiparticle random-phase approximation (QRPA) method \cite{Hinohara}. In these calculation, we use the phase space factor shown in \cite{PhaseSpace}. Therefore, if our experimental sensitivity exceeds that of previous study by approximately an order of magnitude, we can expect either the detection of 2$\nu$2$\beta$ in $^{160}$Gd, or feedback on the NME value giving the shorter half-life prediction through an improved lower limit on the half-life. \\
The PIKACHU project is planned in two experimental phases: Phase~1 aims to search for 2$\beta$ with higher sensitivity than that of the previous study. Phase~2 seeks to either discover 2$\nu$2$\beta$ or constrain the theoretical value of NME by searching for 2$\nu$2$\beta$ with sensitivity comparable to the theoretical prediction mentioned above. We estimated that Phase~1 can be carried out using GAGG crystals with the current level of purity \cite{PIKACHU}. However, the feasibility of Phase~2 requires GAGG crystals with at least an order of magnitude higher purity than the currently used one. In both phases, it is essential to understand the radioactive background that competes with the 2$\beta$ signal, especially for $^{160}$Gd which has a low $Q$-value. This paper describes the development of the background and signal models for the PIKACHU project using Monte-Carlo simulations, and their verification against the measured background spectrum. In addition, we evaluated the Phase~1 sensitivity by fitting the pseudo spectra with the background combined with signal model spectra. 

\section{Monte-Carlo simulation}
\label{GEANT4 simulation}
The background and signal models are essential for evaluating the 2$\beta$ of $^{160}$Gd from the measured energy spectrum. We obtained all model spectra described in this paper, by applying the energy resolution to the energy loss of radiations in GAGG calculated with GEANT4-11.0.2. The Reference PhysicsList library used in GEANT4 is ``Shielding,'' which includes standard and spontaneous nuclear decay physics as well as electromagnetic interactions, including $\alpha$, $\beta$ decays and $\gamma$-rays from nuclear de-excitation. The simulated detector model consists of two physical volumes: the crystal and the light guide. The crystal model is a cylinder with the size of the evaluated crystal. In this paper, we defined a diameter of 54 mm and a height of 52 mm, according to the size of high-purity GAGG \cite{PIKACHU}. Material of the crystal model is GAGG, defined as a mixture of atoms according to the composition formula Gd$_{3}$Ga$_{3}$Al$_{2}$O$_{12}$, with a density of 6.63 g/cm$^{3}$. The light guide model is a trapezoidal cone with a top diameter of 65 mm, a bottom diameter of 50 mm and a height of 100 mm. Its material is defined as G4$\_$PLEXIGLASS prepared in the GEANT4 material database. 

\subsection{Background model}
\label{Background model}
We considered two types of background sources: nuclides in $^{232}$Th, $^{238}$U, and $^{235}$U decay chains contaminated in GAGG, $^{40}$K present in both GAGG and photomultiplier tube (PMT), and $^{208}$Tl present in PMT. 

\subsection*{$^{232}\rm{Th}$, $^{238}\rm{U}$, and $^{235}\rm{U}$ decay chains in GAGG crystal}
\begin{figure}[!h]
    \centering
    \includegraphics[keepaspectratio,scale=0.37]{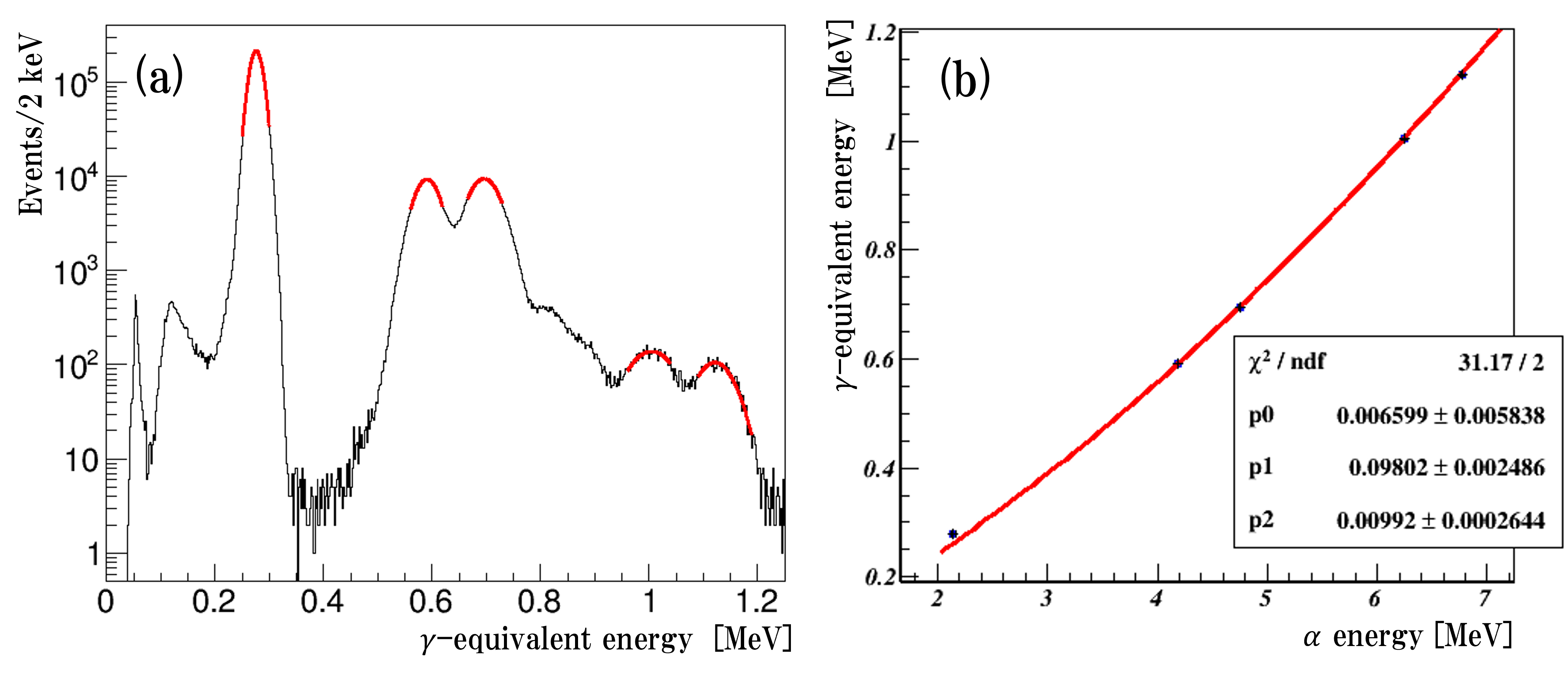}
    \caption{(a) The $\alpha$-ray background spectrum of high-purity GAGG measured for 854.5 hours in Kamioka. The highest peak around 0.25~MeV is due to 2.14 MeV $\alpha$-rays from $^{152}$Gd inherently contained in the GAGG. We evaluated the representative energies after quenching shown by the red Gaussian fits. (b) Quenching effect of the high-purity GAGG on $\alpha$-rays: The blue points represent the $\alpha$-ray energy before quenching (horizontal axis) and the corresponding energy after quenching (vertical axis) obtained from the measured $\alpha$-ray spectrum. These points were fitted with a quadratic function, $y$ = p$_0$ + p$_1$ $x$ + p$_2$ $x^{2}$, as shown by the red line.}
    \label{quenching}
\end{figure}

\begin{figure}[!h]
    \centering
    \includegraphics[keepaspectratio,scale=0.5]{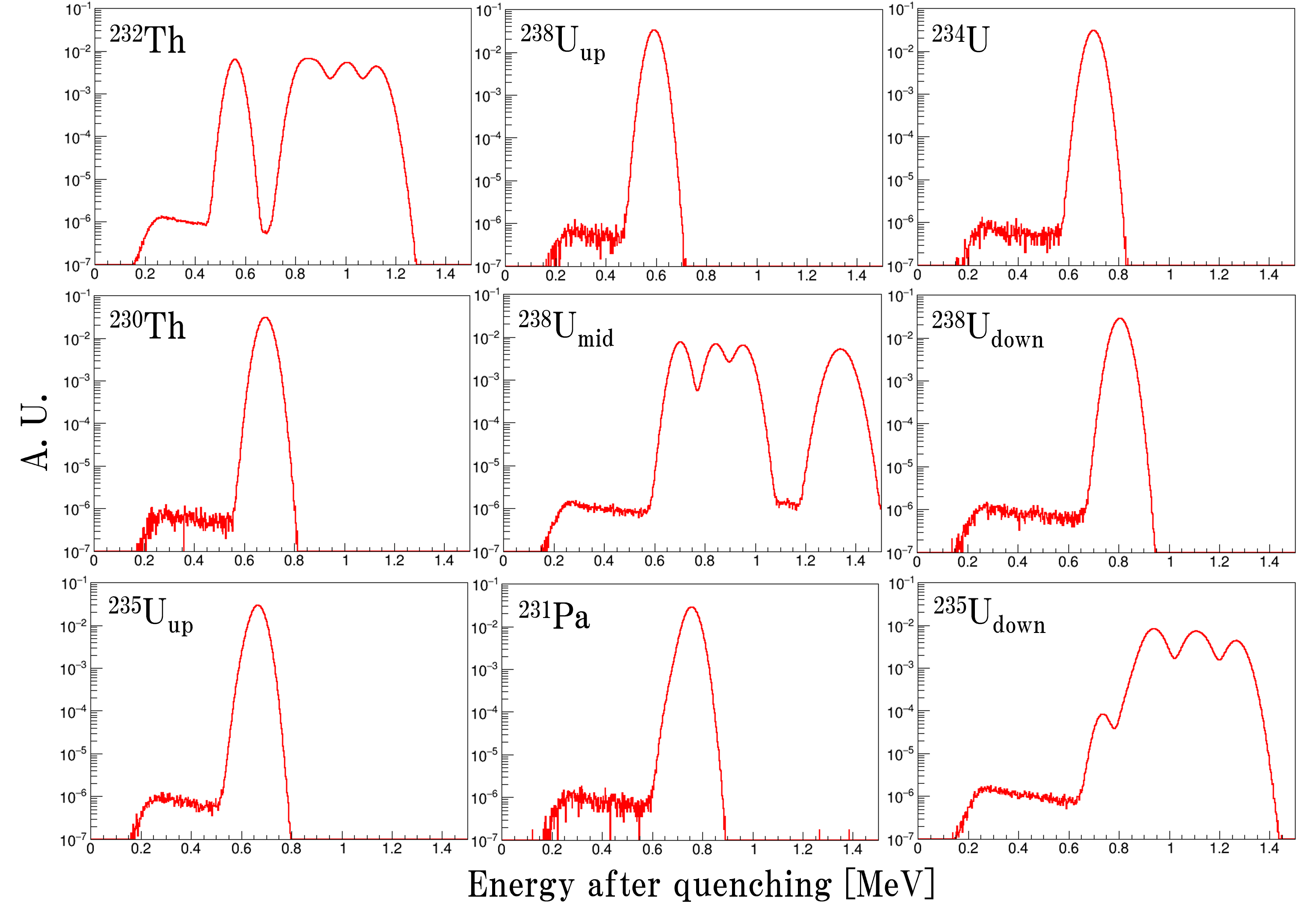}
    \caption{The $\alpha$-ray background model spectra based on GEANT4 simulations. The vertical axes are scaled to normalize the integral of each model spectrum.}
    \label{alpha model}
\end{figure}

\begin{figure}[!h]
    \centering
    \includegraphics[keepaspectratio,scale=0.45]{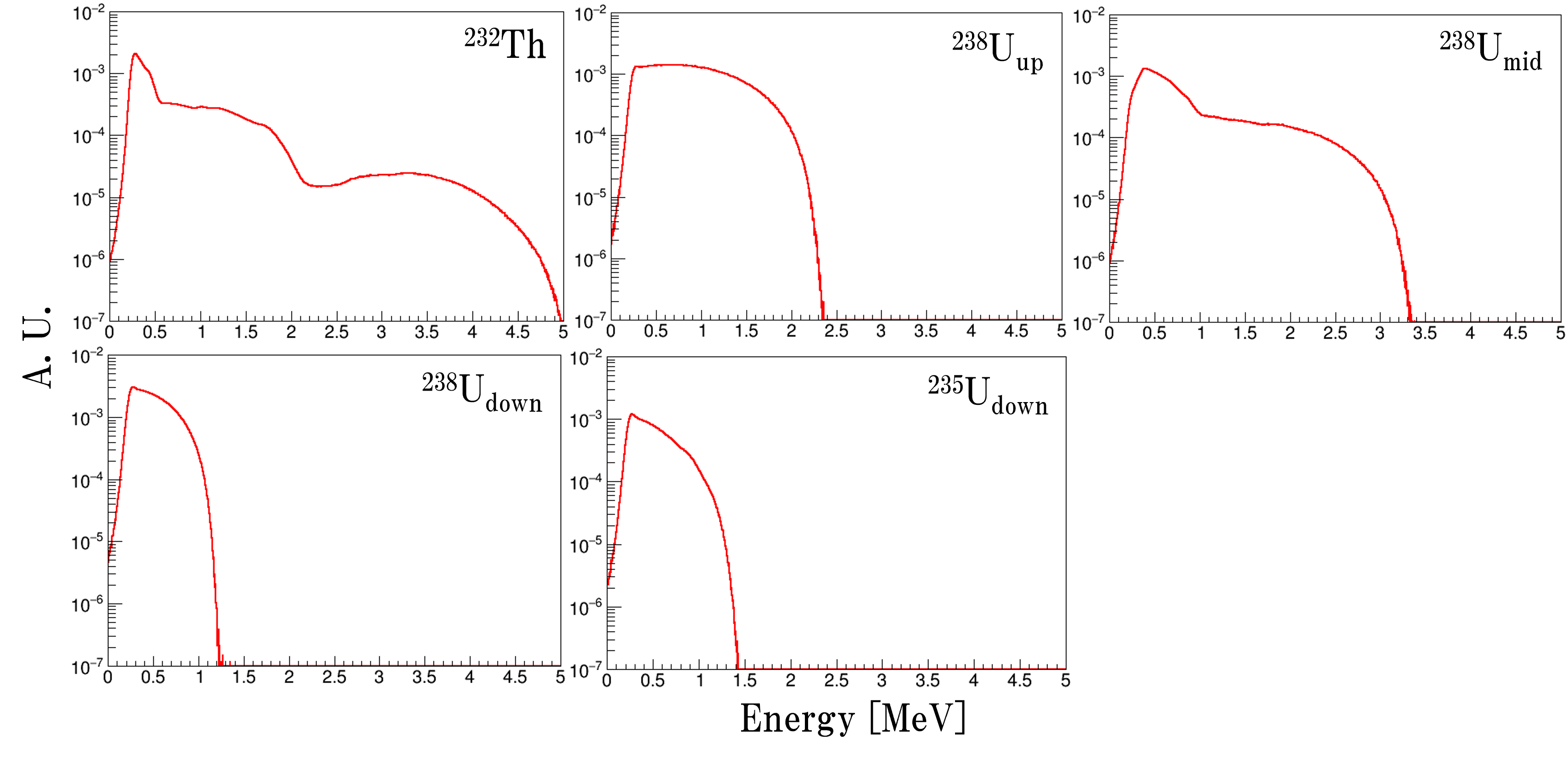}
    \caption{The $\beta$($\gamma$)-ray background model spectra based on GEANT4 simulations. The vertical axes are scaled to normalize the integral of each model spectrum.}
    \label{beta model}
\end{figure}

It has been confirmed through background measurement in Kamioka underground facility that high-purity GAGG contains $^{232}$Th, $^{238}$U, and $^{235}$U decay chains \cite{PIKACHU}. On the other hand, the investigations using a high purity germanium detector demonstrated that the purity of crystal's raw materials is higher than that of high-purity GAGG. This suggests that radioactive impurities may also be introduced from the environment during the crystal growth process. \\
In GEANT4, we generated the stationary parent nuclide for each decay chain, corresponding to $^{232}$Th, $^{238}$U, and $^{235}$U, uniformly within the volume of the crystal model. In this study, each parent nuclide was generated 100 million times. GEANT4 then simulated their chained decays until the parent nuclide reached the most stable nuclide in each chain and the behavior of the resulting radiations in the crystal model. The energy losses of the radiations in the crystal model were recorded, along with the names of the decayed nuclide and the daughter nuclide. However, the recorded energy losses do not reflect the effect of quenching or the detector's energy resolution. These effects are artificially incorporated into analysis as follows.\\
Quenching is the effect in which $\alpha$-rays are detected with energies lower than their original energies. The ratio of the $\alpha$-ray energy calibrated by $\gamma$-rays to the original $\alpha$-ray energy (the $\alpha$/$\gamma$ ratio) is known to depend proportionally on the original $\alpha$-ray energy in GAGG \cite{quenching}. Therefore, we approximated the correlation between the $\alpha$-ray energy before and after quenching using a quadratic function, as shown in Fig. \ref{quenching} (b). The red line represents the quadratic fit function ($y = \mathrm{p_{0}} + \mathrm{p_{1}}\, x + \mathrm{p_{2}}\, x^{2}$) to the blue filled points with error bars. The blue points are obtained by fitting the measured $\alpha$-ray spectrum of high-purity GAGG \cite{PIKACHU} with Gaussian functions (red lines) in Fig. \ref{quenching} (a). Based on the curve in Fig. \ref{quenching} (b), we corrected the energy losses of the $\alpha$-rays calculated by GEANT4 to the $\gamma$-equivalent energies. Then, by applying the detector's resolution to the corrected energy losses, we reproduced the measured $\alpha$-ray spectrum. The resolution was measured using $\gamma$-ray sources and approximated as a function of $\gamma$-ray energy with an offset of 0.20~MeV, as $\sigma/E = \frac{1.9}{\sqrt{E - 0.20}} + 1.1$. For this reason, we applied the resolution to $\gamma$-equivalent energies above 0.20~MeV. Therefore, in all model spectra presented in present paper, only permeation events due to resolution can be confirmed below 0.20~MeV. However, this effect of the offset is negligible because it is sufficiently lower than the energy region of interest in our analysis.\\
Radioactive nuclides are present in GAGG in a state of radioactive equilibrium in relation to half-lives as shown in Table \ref{radioative equilibrium}. We classified the $^{238}$U decay chain into five sections: $^{238}$U$_{\rm{up}}$, $^{234}$U, $^{230}$Th, $^{238}$U$_{\rm{mid}}$, and $^{238}$U$_{\rm{down}}$, with $^{238}$U, $^{234}$U,  $^{230}$Th, $^{226}$Ra, and $^{210}$Pb as the parent nuclides, respectively. $^{235}$U decay chain was divided into $^{235}$U$_{\rm{up}}$, $^{231}$Pa, and $^{235}$U$_{\rm{down}}$ with $^{235}$U, $^{231}$Pa, and $^{227}$Ac as the parent nuclides, respectively. We developed the $\alpha$-ray and $\beta$($\gamma$)-ray background model spectra for each equilibrium section, as shown in Figs. \ref{alpha model} and \ref{beta model}. Each spectrum in Fig. \ref{alpha model} has an almost flat component that extends toward the low energy side up to 0.20~MeV, due to events involving the escape of $\alpha$-ray or other secondary particles.

\renewcommand{\arraystretch}{1.5}
\begin{table}[!h]
\caption{Radioactive equilibrium for $^{232}$Th, $^{238}$U, and $^{235}$U decay chains assumed in our experiment.}
\label{radioative equilibrium}
\centering
\begin{tabular}{llll}\hline
Chain & Section name & Parent nucleus ($T_{1/2}$ [Yr]) & Last nucleus \\ [2ex]\hline
$^{232}$Th & $^{232}$Th & $^{232}$Th ($1.4\times10^{10}$)& $^{208}$Pb\\\hline
\multirow{5}{*}{$^{238}$U} & $^{238}$U$_{\rm{up}}$ & $^{238}$U ($4.5\times10^{9}$) & $^{234}$U \\
 & $^{234}$U & $^{234}$U ($2.5\times10^{5}$) & $^{230}$Th \\ 
 & $^{230}$Th & $^{230}$Th ($7.5\times10^{4}$) & $^{226}$Ra \\
 & $^{238}$U$_{\rm{mid}}$ & $^{226}$Ra ($1.6\times10^{3}$) &  $^{210}$Pb\\
 & $^{238}$U$_{\rm{down}}$ & $^{210}$Pb (22.3) & $^{206}$Pb \\\hline
\multirow{3}{*}{$^{235}$U} & $^{235}$U$_{\rm{up}}$ & $^{235}$U ($7.0\times10^{8}$) & $^{231}$Pa\\
 & $^{231}$Pa & $^{231}$Pa ($3.3\times10^{4}$) & $^{227}$Ac \\
 & $^{235}$U$_{\rm{down}}$ & $^{227}$Ac (21.8) & $^{207}$Pb \\ \hline 
\end{tabular}
\end{table}

\subsection*{$^{40}\rm{K}$ in GAGG and PMT}
\begin{figure}[!h]
    \centering
    \includegraphics[keepaspectratio,scale=0.55]{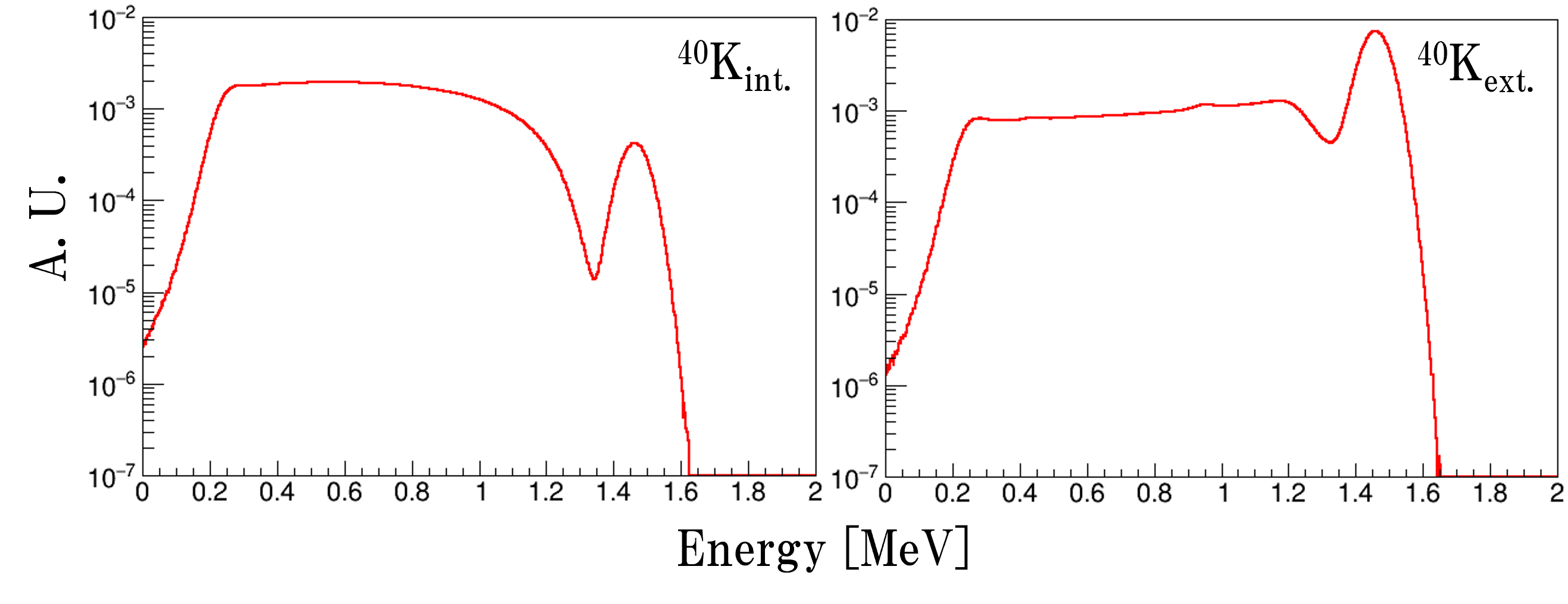}
    \caption{Background model spectrum for $^{40}$K$_{\rm{int.}}$ and $^{40}$K$_{\rm{ext.}}$ based on GEANT4 simulations. The vertical axes are scaled to normalize the integral of each model spectrum.}
    \label{K40 model}
\end{figure}

$^{40}$K is an environmental isotope with a natural abundance of 0.0117$\%$. The most common disintegration mode is the beta-minus decay to the ground state of $^{40}$Ca, with a $Q_{\beta^{-}}$ of 1.31~MeV, occurring with a branching ratio of 89.25$\%$. The next dominant mode is the electron capture, which converts $^{40}$K to the excited state of $^{40}$Ar, followed by the de-excitation to the ground state with the emission of a 1.46~MeV $\gamma$-ray, occurring with a branching ratio of 10.55$\%$. There are also other modes with branching ratios on the order of less than 0.1$\%$ \cite{40K}. Based on the above, the energy spectrum of $^{40}$K has a photoelectric peak at 1.46~MeV and a continuous beta spectrum up to around 1.31~MeV. Therefore, the radiation emitted from $^{40}$K can contribute to the background in the 2$\nu$2$\beta$ search because its spectrum is continuous up to around 1.73~MeV. In contrast, $^{40}$K would not contribute significantly to the background for the search of 0$\nu$2$\beta$, which is expected to produce a peak spectrum at 1.73~MeV.\\
$^{40}$K is present in both PMT ($^{40}$K$_{\rm{ext.}}$) and GAGG ($^{40}$K$_{\rm{int.}}$). The $^{40}$K$_{\rm{ext.}}$ emits 1.46~MeV $\gamma$-ray, which is detected as a background when it enters and deposits the energy in GAGG. Assuming that $^{40}$K is contained on the photocathode surface of the PMT, we simulated spherically symmetric emissions of 1.46~MeV $\gamma$-rays from uniform positions on the 50 mm-diameter surface of the light guide model, which is the joint surface with the PMT. For $^{40}$K$_{\rm{int.}}$, we generated stationary $^{40}$K uniformly in the crystal model and simulated the aforementioned disintegration scheme.\\
Consequently, we obtained the background model spectra for $^{40}$K as shown in Fig. \ref{K40 model}. The energy resolution was applied as described in the previous section. 

\subsection*{$^{208}\rm{Tl}$ in PMT}
\begin{figure}[!h]
    \centering
    \includegraphics[keepaspectratio,scale=0.5]{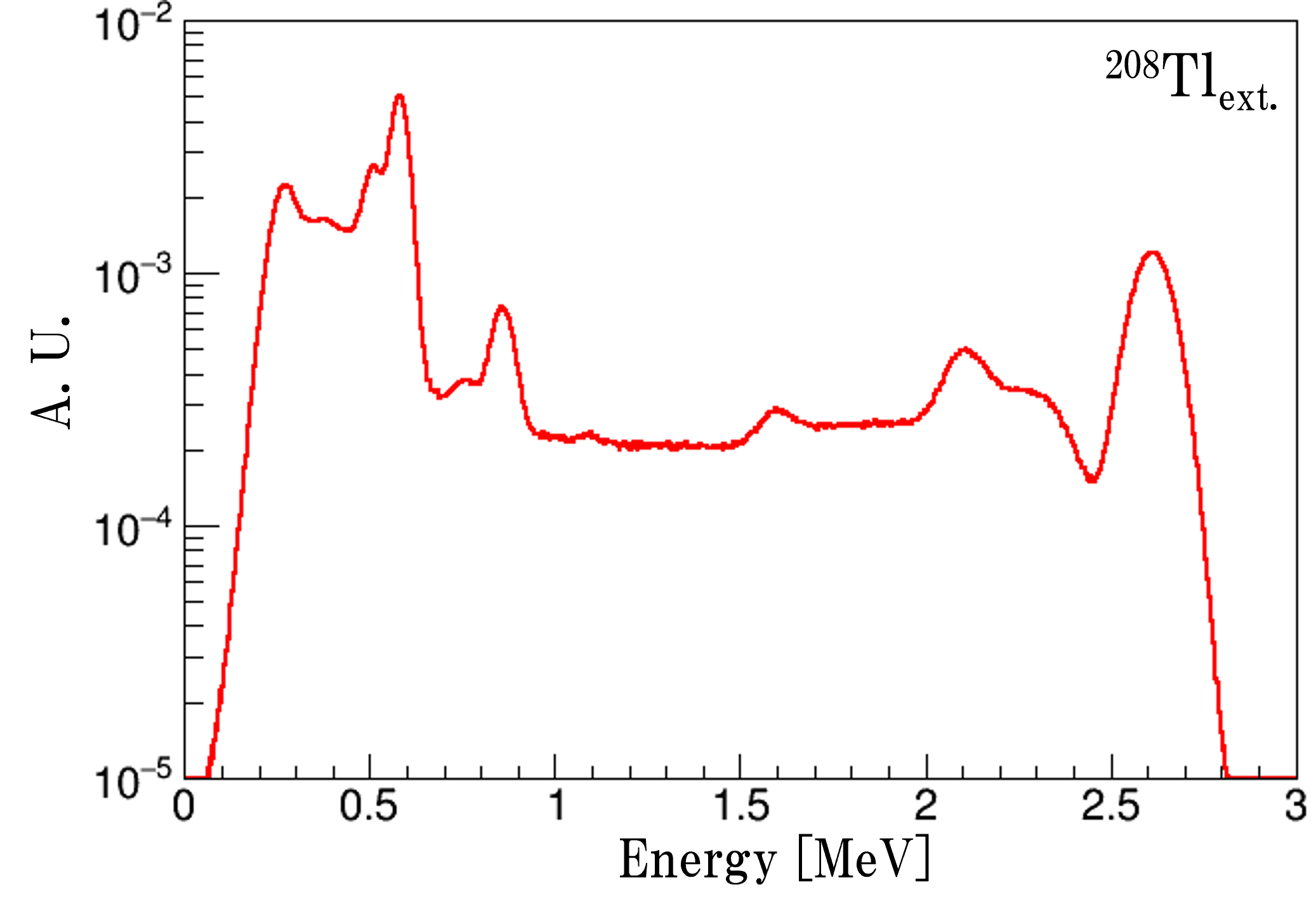}
    \caption{Background model spectrum for $^{208}$Tl$_{\rm{ext.}}$ based on GEANT4 simulations. The vertical axes are scaled to normalize the integral of the model spectrum. The main $\gamma$-rays (2.614~MeV, 860.5~keV, 583.2~keV, and 540.7~keV) are identified as photoelectric peaks. The small peaks around 1.6~MeV and 2.1 MeV are attributed to single and double escape events resulting from 2.614~MeV $\gamma$-ray.}    
    \label{Tl208 model}
\end{figure}

$^{208}$Tl is a member of the $^{232}$Th decay chain and undergoes a beta-minus decay to various excited levels of $^{208}$Pb with a $Q_{\beta^{-}}$ of 5.00~MeV \cite{208Tl}. When $^{208}$Tl decays in our PMT, the 2.614~MeV, 860.5~keV, 583.2~keV, and 540.7~keV $\gamma$-ray emitted mainly from de-excitated $^{208}$Pb can be the $\beta$($\gamma$)-ray background. In order to simulate the energy deposit of these $\gamma$-rays from the PMT ($^{208}$Tl$_{\rm{ext.}}$), we generated the stationary $^{208}$Tl in uniform positions on the 50 mm-diameter surface of the light guide model. The model spectrum of $^{208}$Tl$_{\rm{ext.}}$ is shown in Fig. \ref{Tl208 model}.

\subsection{Signal model}\label{Signal model}
\begin{figure}[!h]
    \centering
    \includegraphics[keepaspectratio,scale=0.5]{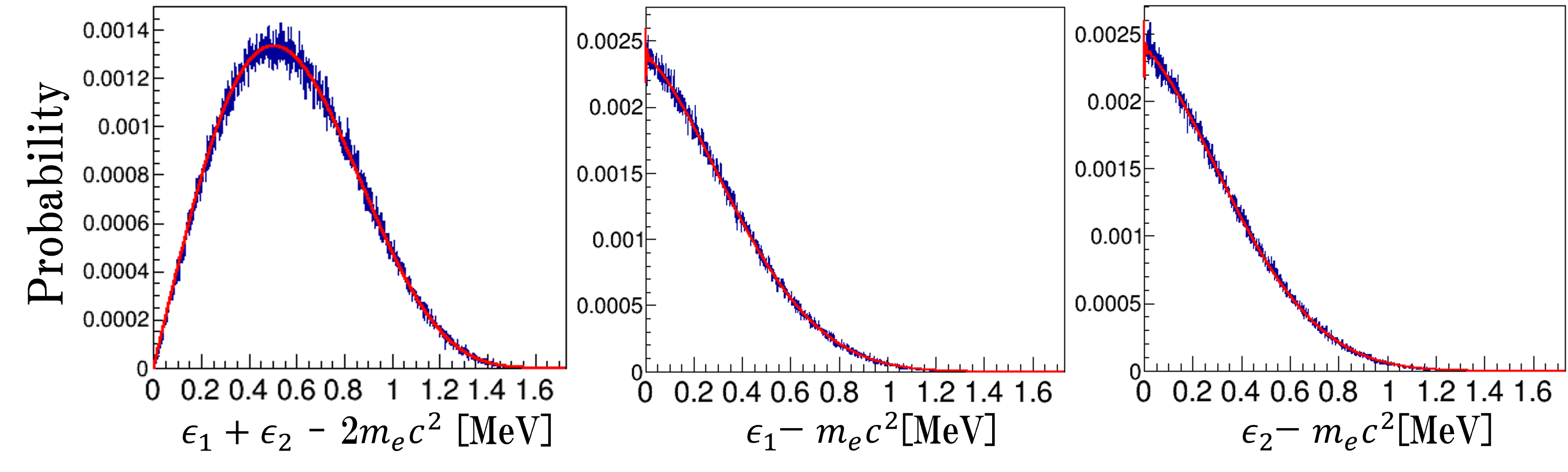}
    \caption{Energy distribution of the summed (Left) and individual electrons (Center and Right) emitted in the 2$\nu$2$\beta$ of $^{160}$Gd. The red lines represent the theoretical probability distributions \cite{PhaseSpace}, while the blue histograms show the distributions calculated using the inverse function method.}
    \label{2b energy}
\end{figure}

\begin{figure}[!h]
    \centering
    \includegraphics[keepaspectratio,scale=0.45]{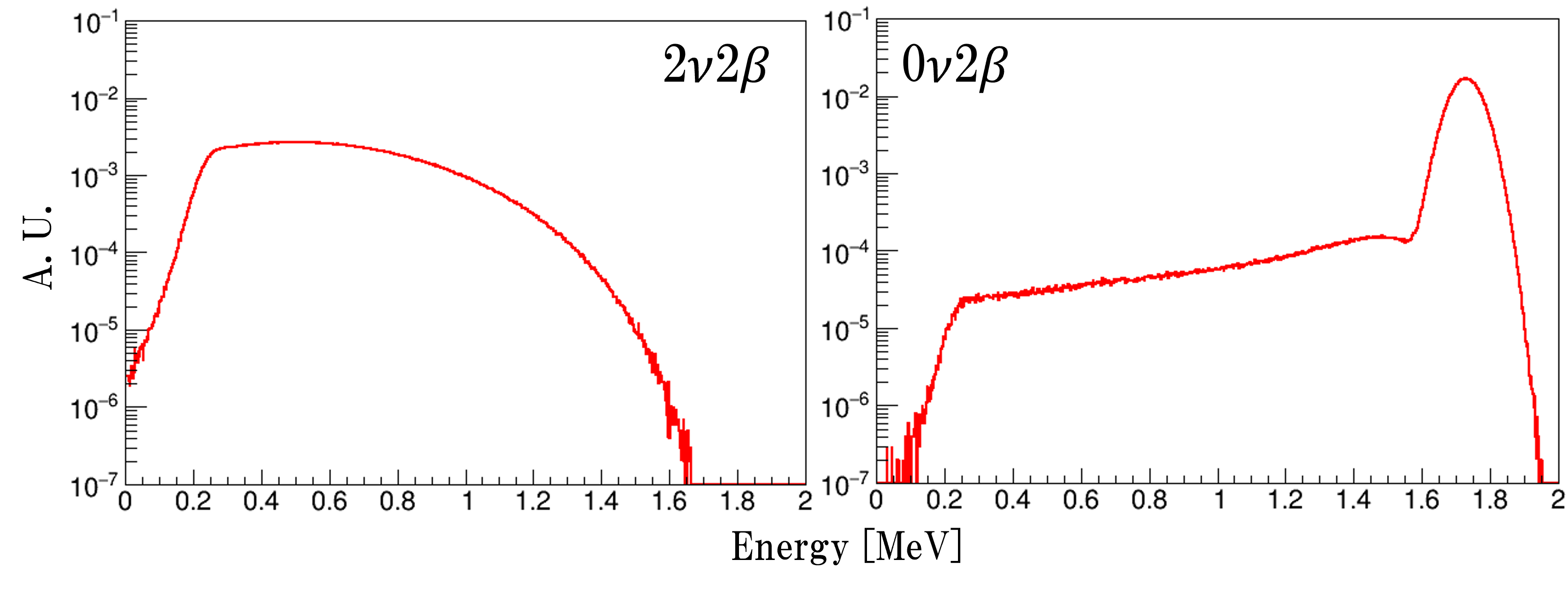}
    \caption{The 2$\nu$2$\beta$ and 0$\nu$2$\beta$ model spectra based on GEANT4 simulations. The vertical axes are scaled to normalize the integral of each model spectrum.}
    \label{2b model}
\end{figure}

The 2$\beta$ in GAGG can be simulated by generating two electrons from the same position, with the energies governed by theoretical probability distributions. The kinematics of the emitted electrons is described in terms of phase space factors. Reference \cite{PhaseSpace} calculated phase space factors for various 2$\beta$ candidates across a broad range of mass numbers, from $^{48}$Ca to $^{238}$U. Reference \cite{PhaseSpace} also provides a list of combinations of each electron energy ($\epsilon_{1, 2}$) minus the rest mass energy ($m_{e}c^{2}$) emitted in the 2$\nu$2$\beta$, denoted as ($\epsilon_{1}-m_{e}c^{2}$, $\epsilon_{2}-m_{e}c^{2}$), along with their corresponding probabilities of occurrence, given at intervals of 1~keV. By summing up the probabilities in the list specific to $^{160}$Gd, we produced a discrete cumulative distribution function (CDF) of ($\epsilon_{1}-m_{e}c^{2}$, $\epsilon_{2}-m_{e}c^{2}$). The CDF can then be used to determine ($\epsilon_{1}-m_{e}c^{2}$, $\epsilon_{2}-m_{e}c^{2}$) based on uniform random numbers between 0 and 1, like an inversion method \cite{inversion}. \\
The blue histograms in Fig. \ref{2b energy} represent the energy distributions of the summed electrons, $\epsilon_{1} + \epsilon_{2} -2m_{e}c^{2}$ (Left) and each individual electron, $\epsilon_{1}-m_{e}c^{2}$ (Center) and $\epsilon_{2}-m_{e}c^{2}$ (Right), calculated 700,000 times using the CDF. Figure \ref{2b energy} shows that the CDF can determine the energies of the two electrons in accordance with the theoretical distributions represented by the red lines. In GEANT4, we defined two primary electrons emitted with energies $\epsilon_{1}$ and $\epsilon_{2}$ obtained from the CDF, emitted in random directions from a random position in the crystal model. The resulting 2$\nu$2$\beta$ model spectrum is shown on the left in Fig. \ref{2b model}.\\
The theoretical energy distribution for one of the two electrons emitted in 0$\nu$2$\beta$ is also given in Ref. \cite{PhaseSpace}. In this case, we determined only $\epsilon_{1}$ from the CDF, and calculated $\epsilon_{2}$ by subtracting $\epsilon_{1}$ from the $Q$-value. The 0$\nu$2$\beta$ model spectrum is shown on the right in Fig. \ref{2b model}. It has a peak at the $Q$-value (1.73~MeV) and also has a nearly flat tail component below 1.5~MeV due to the escape of electrons from the crystal. The energy resolution was applied as described in Section \ref{Background model}. 

\section{Sensitivity study using the background and signal models}
\label{Sensitivity study}
\begin{figure}[h]
    \centering
    \begin{subfigure}[b]{0.45\textwidth}
        \centering
        \includegraphics[width=\linewidth]{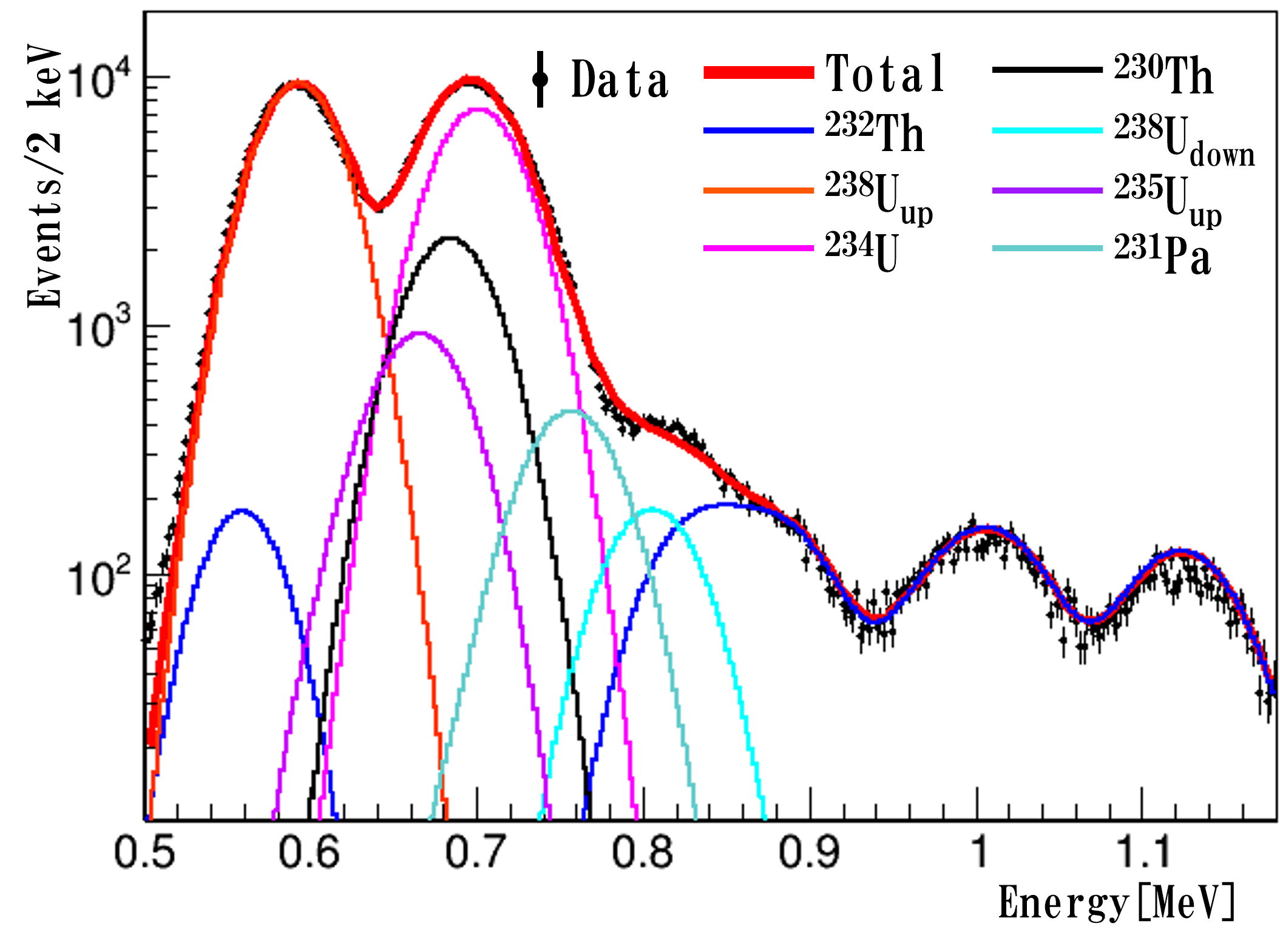}
        \caption{The $\alpha$-ray spectrum. $\chi^{2}$/ndf is 3624.4/331.}
        \label{alpha bestfit}
    \end{subfigure}
    \hfill
    \begin{subfigure}[b]{0.45\textwidth}
        \centering
        \includegraphics[width=\linewidth]{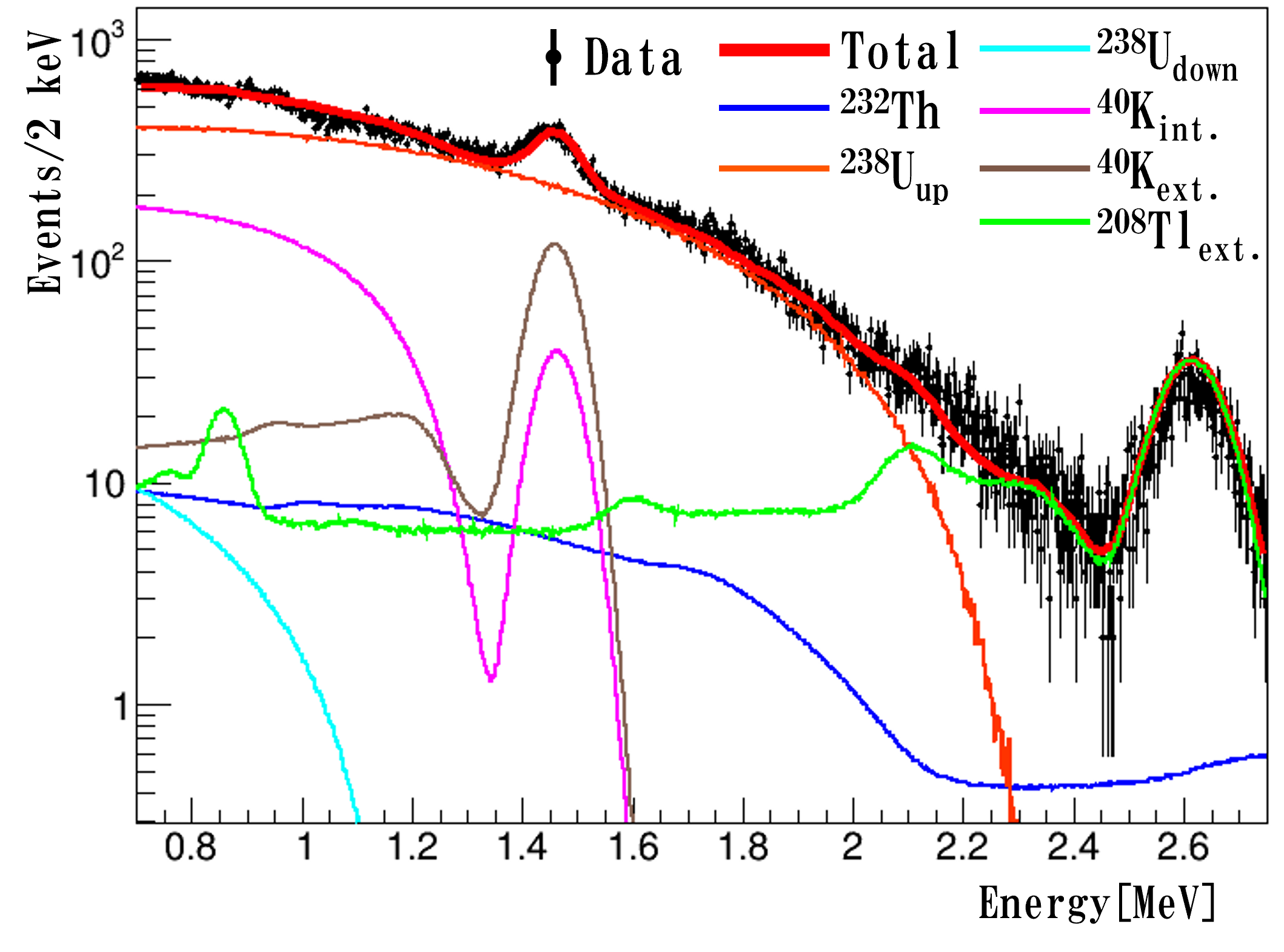}
        \caption{The $\beta$($\gamma$)-ray spectrum. $\chi^{2}$/ndf is 1799.9/1022.}
        \label{beta bestfit}
    \end{subfigure}
    \caption{Measured $\alpha$-ray and $\beta$($\gamma$)-ray spectra fitted with their respective model spectra. The data are shown as black points. The total model spectrum and each background source are represented by thick red lines and thin solid lines in different colors, respectively.}
    \label{bestfit}
\end{figure}

The purpose of Phase~1 of the PIKACHU project is to update the lower limits of $T^{2\nu(0\nu)}_{1/2}$ for $^{160}$Gd. Based on the performance evaluation of high-purity GAGG \cite{PIKACHU}, Phase~1 experiment is planned to last for two years, using two large GAGG single crystals (with dimensions specified in Section \ref{Introduction}) with a purity equivalent to that of high-purity GAGG. In this section, we estimate the sensitivity that Phase~1 can achieve. \\
The background models developed as described in Section \ref{Background model} reproduced the measured background by fitting as shown in Fig. \ref{bestfit}. The black points represent high-purity GAGG's $\alpha$-ray (Fig. \ref{alpha bestfit}) and $\beta$($\gamma$)-ray (Fig. \ref{beta bestfit}) background spectrum measured for 854.5 hours in Kamioka. This measurement was made using a setup consisting of high-purity GAGG, an acrylic light guide, and a PMT (R6231-100 manufactured by Hamamatsu Photonics \cite{r6231}). The thick red lines represent the total of background model spectra fitted to the data. The fitting ranges are 0.5 $\sim$ 1.18 MeV and 0.7 $\sim$ 2.75 MeV for $\alpha$-ray and $\beta$($\gamma$)-ray background spectrum, respectively. While the thin lines in different colors represent the breakdown of the background components, as indicated in the legends. Because the $\alpha$-ray backgrounds arise solely from $\alpha$ decays in the U/Th chains present in GAGG, the fitting parameters for $\alpha$-ray spectrum are the amount of U/Th impurities listed in Table \ref{radioative equilibrium}. As for $\beta$($\gamma$)-ray spectrum, we fixed the model spectra of the U/Th impurities to the amounts obtained from the $\alpha$-ray spectrum fitting, and fitted additionally the model of $^{40}$K$_{int.}$, $^{40}$K$_{ext.}$, and $^{208}$Tl$_{ext.}$. \\
To evaluate the sensitivity in Phase~1, we scaled the best-fit background model spectrum (thick red lines in Fig. \ref{bestfit}) to the Phase~1 exposure of 1.43 kg $\cdot$ year, and determined the number of events for each bin according to the Poisson distribution, with averages set to the values of the scaled best-fit at corresponding bins. The spectrum generated pseudo-randomly in this way is referred to as the pseudo spectrum in the present paper. We generated 10,000 sets of the $\alpha$-ray and $\beta$($\gamma$)-ray pseudo background spectrum, and evaluated the sensitivity using two different methods, as described below. 

\subsection*{Method 1}
\begin{figure}[!h]
    \centering
    \begin{subfigure}[b]{0.45\textwidth}
        \centering
        \includegraphics[width=\linewidth]{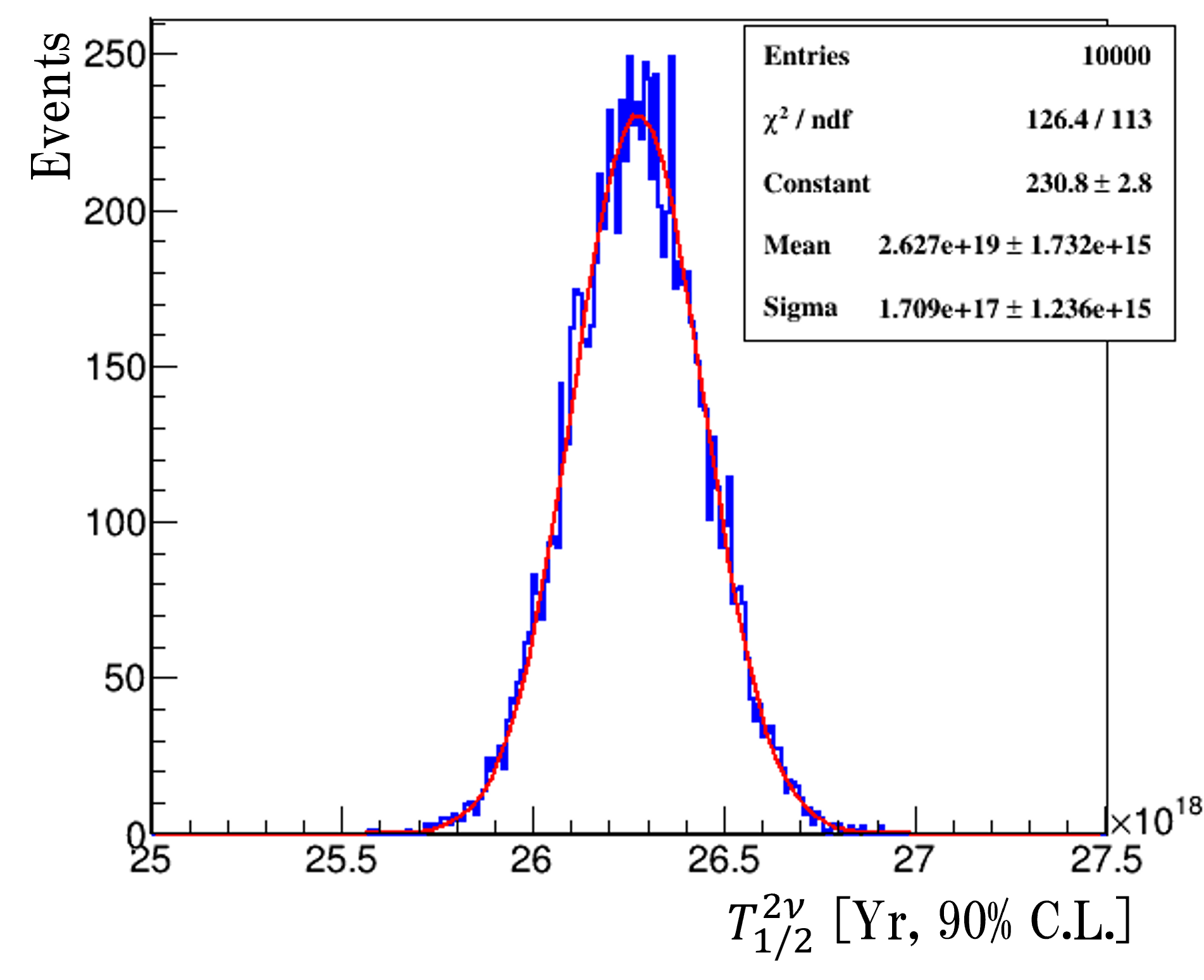}
        \caption{For 2$\nu$2$\beta$. The mean of lower limits of $T^{2\nu}_{1/2}$ is 2.63~$\times$~10$^{19}$~years.}
        \label{2v limit}
    \end{subfigure}
    \hfill
    \begin{subfigure}[b]{0.45\textwidth}
        \centering
        \includegraphics[width=\linewidth]{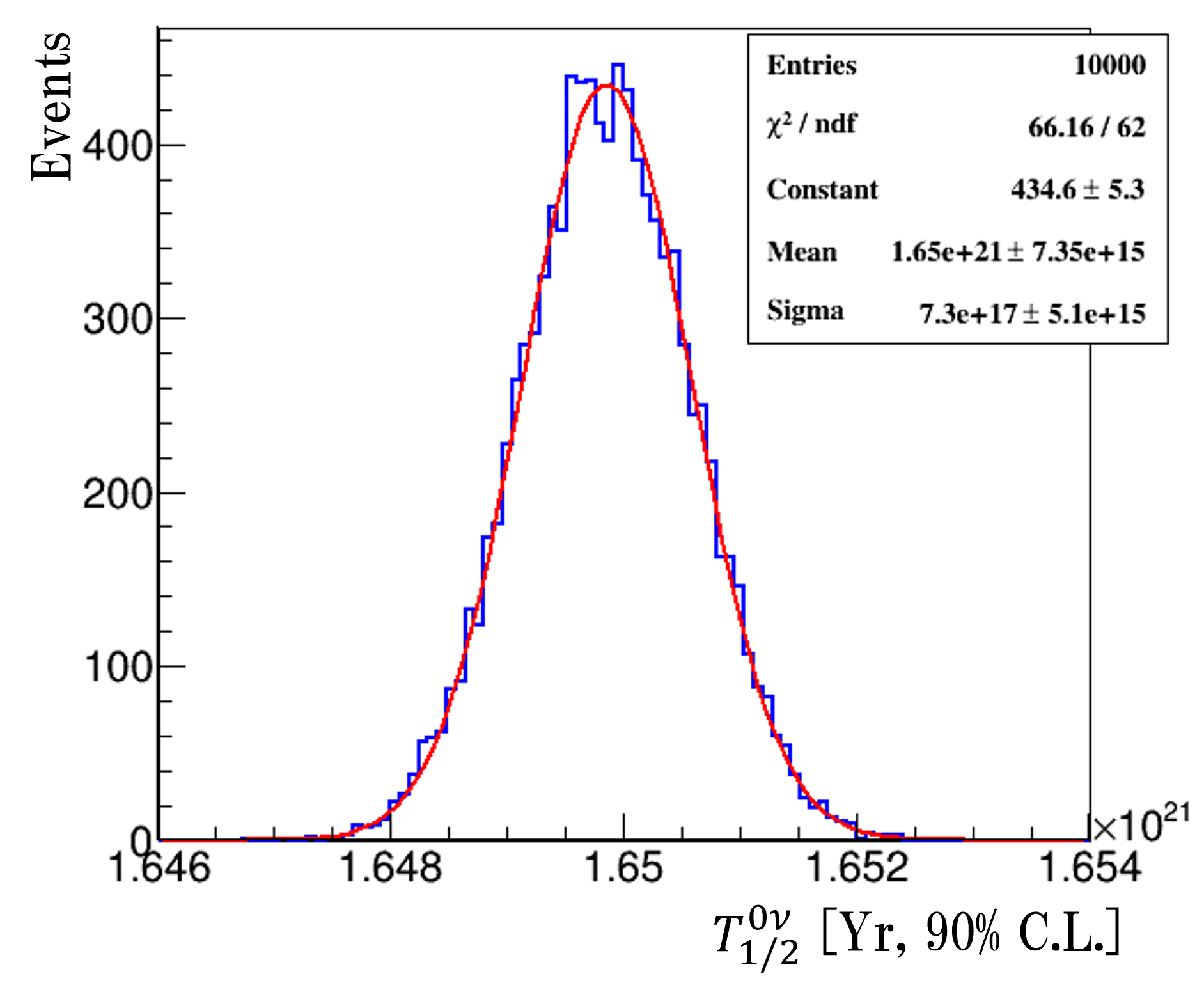}
        \caption{For 0$\nu$2$\beta$. The mean of lower limits of $T^{0\nu}_{1/2}$ is 1.65~$\times$~10$^{21}$~years.}
        \label{0v limit}
    \end{subfigure}
    \caption{The distributions of lower limits of half-lives at 90$\%$ C.L. estimated by Method~1. The red lines represent the Gaussian fitting results.}
    \label{error limit}
\end{figure}

In this method, we fitted all the pseudo spectra with the background models combined with the 2$\nu$2$\beta$ (0$\nu$2$\beta$) signal model. Except for the addition of the signal model to the fitting parameters of the $\beta$($\gamma$)-ray pseudo spectrum, the fitting conditions are the same as those in the analysis shown in Fig. \ref{bestfit}. We regarded the error in the 2$\nu$2$\beta$ (0$\nu$2$\beta$) signal parameter as the upper limit for the number of observed signals ($N^{2\nu(0\nu)}_{\rm{obs}}$). We then calculated the lower limits of $T^{2\nu(0\nu)}_{1/2}$ using

\begin{equation}\label{sens}
\begin{split}
T^{2\nu(0\nu)}_{1/2} = (\rm ln \ 2) \it N_{A}\frac{\epsilon M a}{W N^{\rm{2\nu(0\nu)}}_{\rm{obs}}}t,
\end{split}
\end{equation}

\noindent where the Avogadro constant ($N_{A}$) is 6.02 $\times$ 10$^{23}$ atoms/mol, the natural abundance of $^{160}$Gd ($a$) is 0.219, the atomic mass of $^{160}$Gd ($W$) is 0.157 kg/mol, the detection efficiency ($\epsilon$) is assumed to be 1.00, the amount of Gd contained in two GAGG crystals ($M$) is 3.25 kg, and the search period ($t$) is 2.00 years. Figures \ref{2v limit} and \ref{0v limit} show the distributions of the lower limits of $T^{2\nu}_{1/2}$ and $T^{0\nu}_{1/2}$ at 90$\%$ C.L., respectively, calculated for all the pseudo spectra. As shown in Fig. \ref{error limit}, the lower limits of $T^{2\nu}_{1/2}$ and $T^{0\nu}_{1/2}$ expected in Phase~1 are 2.63 $\times$ 10$^{19}$ years and 1.65 $\times$ 10$^{21}$ years, respectively.

\subsection*{Method 2}
\begin{figure}[!h]
    \centering
    \begin{subfigure}[b]{0.47\textwidth}
        \centering
        \includegraphics[width=\linewidth]{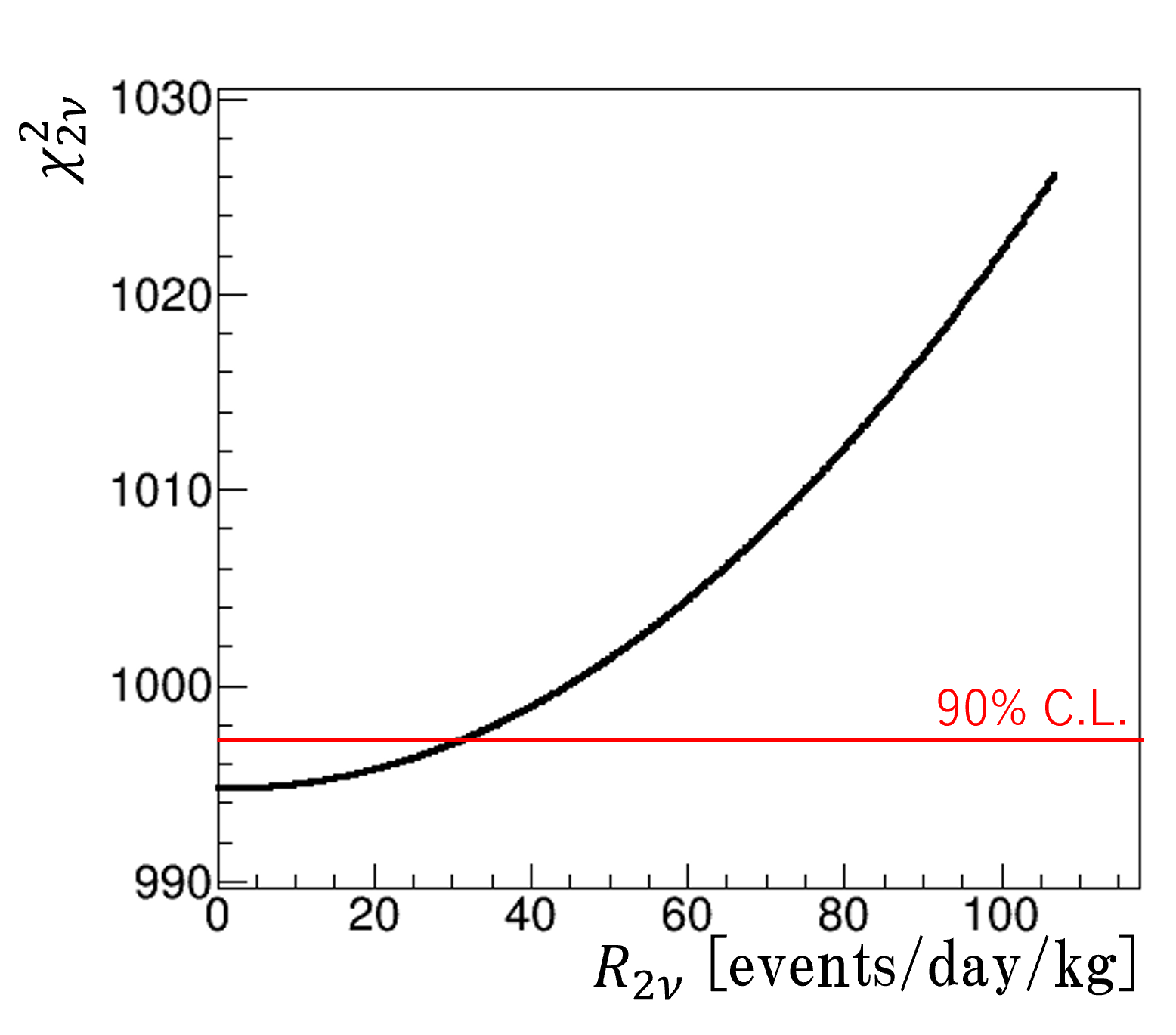}
        \caption{$\chi_{2\nu}^{2}$ as a function of 2$\nu$2$\beta$ rate. The upper limit of 2$\nu$2$\beta$ rate is 31.8 events/day/kg at 90$\%$ C.L.}
        \label{2v chi2}
    \end{subfigure}
    \hfill
    \begin{subfigure}[b]{0.47\textwidth}
        \centering
        \includegraphics[width=\linewidth]{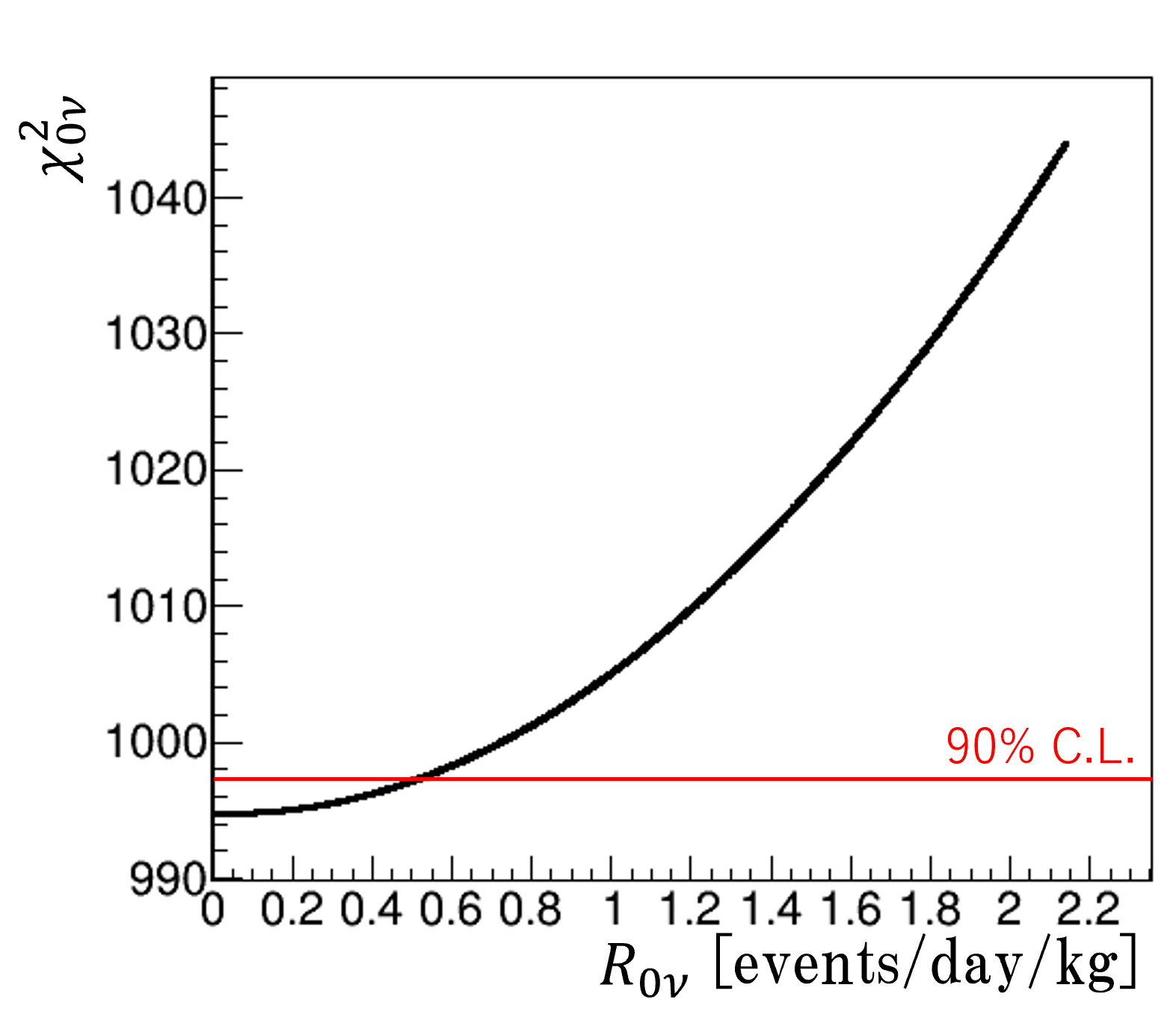}
        \caption{$\chi_{0\nu}^{2}$ as a function of 0$\nu$2$\beta$ rate. The upper limit of 0$\nu$2$\beta$ rate is 0.51 events/day/kg at 90$\%$ C.L.}
        \label{0v chi2}
    \end{subfigure}
    \caption{The $Goodness$-$of$-$Fit$ curves as a function of the event rate for 2$\nu$2$\beta$ and 0$\nu$2$\beta$ in a particular $\beta$($\gamma$)-ray pseudo spectrum. The red lines represent the $\chi_{2\nu}^{2}$ and $\chi_{0\nu}^{2}$ corresponding to the 90$\%$ C.L.}
    \label{2brate vs chi2}
\end{figure}

In this method, we investigated the $Goodness$-$of$-$Fit$ ($\chi_{2\nu(0\nu)}^{2}$) for the $\beta$($\gamma$)-ray pseudo spectrum by increasing 2$\nu$2$\beta$ (0$\nu$2$\beta$) signal rate from zero. The amount of U/Th impurities in GAGG was determined from each $\alpha$-ray pseudo spectrum as before. $^{40}$K$_{\rm{int.}}$, $^{40}$K$_{\rm{ext.}}$, and $^{208}$Tl$_{\rm{ext.}}$ are fitting parameters associated with 2$\nu$2$\beta$ (0$\nu$2$\beta$) signal rate. The $\chi_{2\nu(0\nu)}^{2}$ is given by

\begin{equation}\label{chi2}
\begin{split}
\chi_{2\nu(0\nu)}^{2} = 2 \sum_{i=1}^N \left[ \mu^{2\nu(0\nu)}_{i} - n_{i} + n_{i} \ln \frac{n_{i}}{\mu^{2\nu(0\nu)}_{i}} \right],
\end{split}
\end{equation}

\noindent where $\mu^{2\nu(0\nu)}_{i}$ and $n_{i}$ represent the values of the total model included 2$\nu$2$\beta$ (0$\nu$2$\beta$) signal and data in the $i$-th bin \cite{BakerCousins}. The final term of the right-hand side in Eq. (\ref{chi2}) is zero for bins with no data ($n_{i}$ = 0). Figure \ref{2v chi2} (\ref{0v chi2}) shows the $\chi_{2\nu(0\nu)}^{2}$ as a function of 2$\nu$2$\beta$ (0$\nu$2$\beta$) signal rate, $R_{2\nu(0\nu)}$ [events/day/kg], for a particular $\beta$($\gamma$)-ray pseudo spectrum. The $\chi_{2\nu(0\nu)}^{2}$ provides the Probability at a certain signal rate $R'_{2\nu(0\nu)}$ by

\begin{equation}\label{probability}
\begin{split}
{\rm{Probability}} = \frac{\int_{0}^{R'_{2\nu(0\nu)}} \exp(-\chi_{2\nu(0\nu)}^2)\;dR_{2\nu(0\nu)}}{\int_{0}^{\infty} \exp(-\chi_{2\nu(0\nu)}^2)\;dR_{2\nu(0\nu)}}
\end{split}
\end{equation}

\begin{figure}[!h]
    \centering
    \begin{subfigure}[b]{0.47\textwidth}
        \centering
        \includegraphics[width=\linewidth]{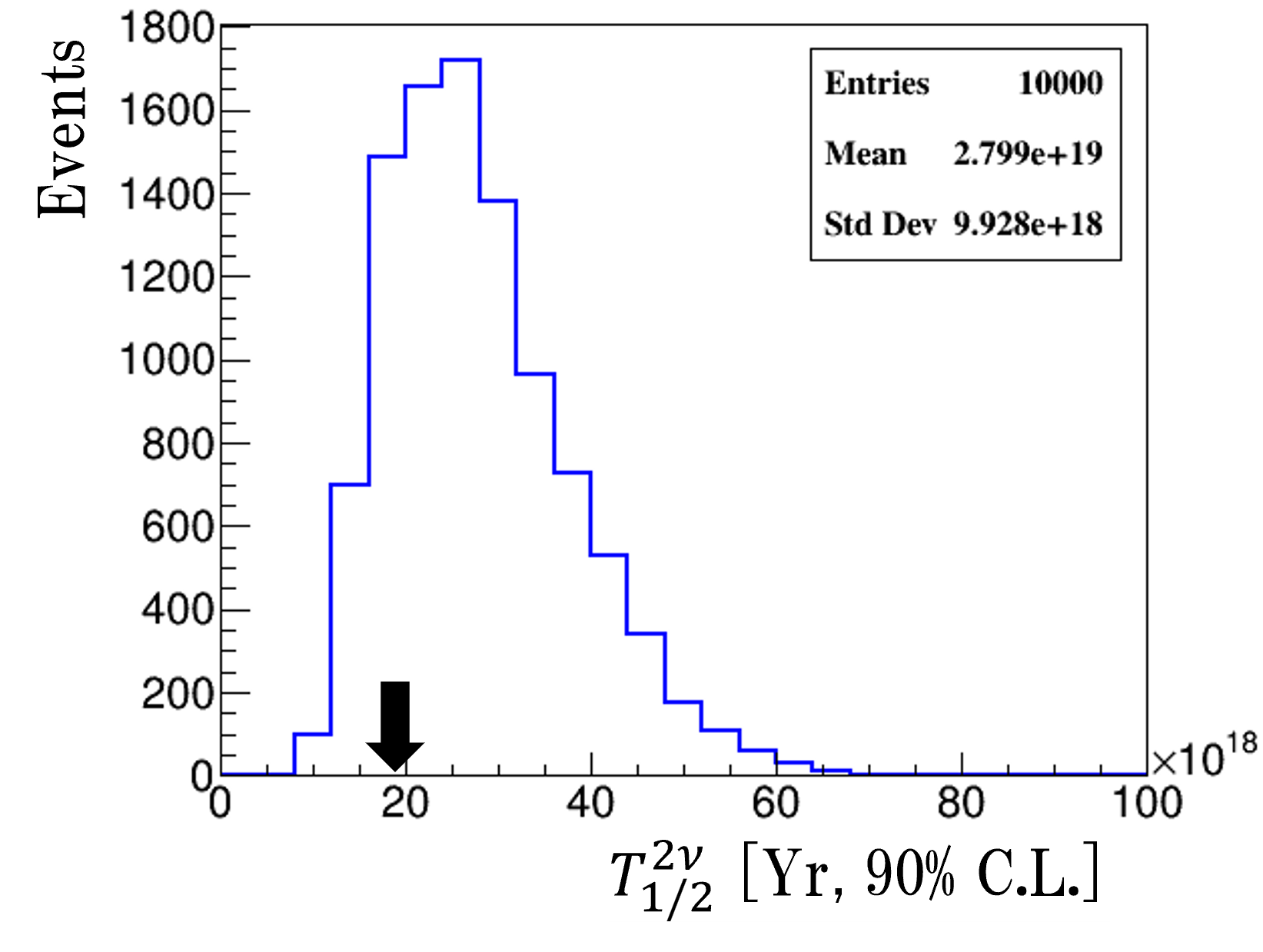}
        \caption{For 2$\nu$2$\beta$. The median is 2.64 $\times$ 10$^{19}$ years.}
        \label{chi2 2v limit}
    \end{subfigure}
    \hfill
    \begin{subfigure}[b]{0.47\textwidth}
        \centering
        \includegraphics[width=\linewidth]{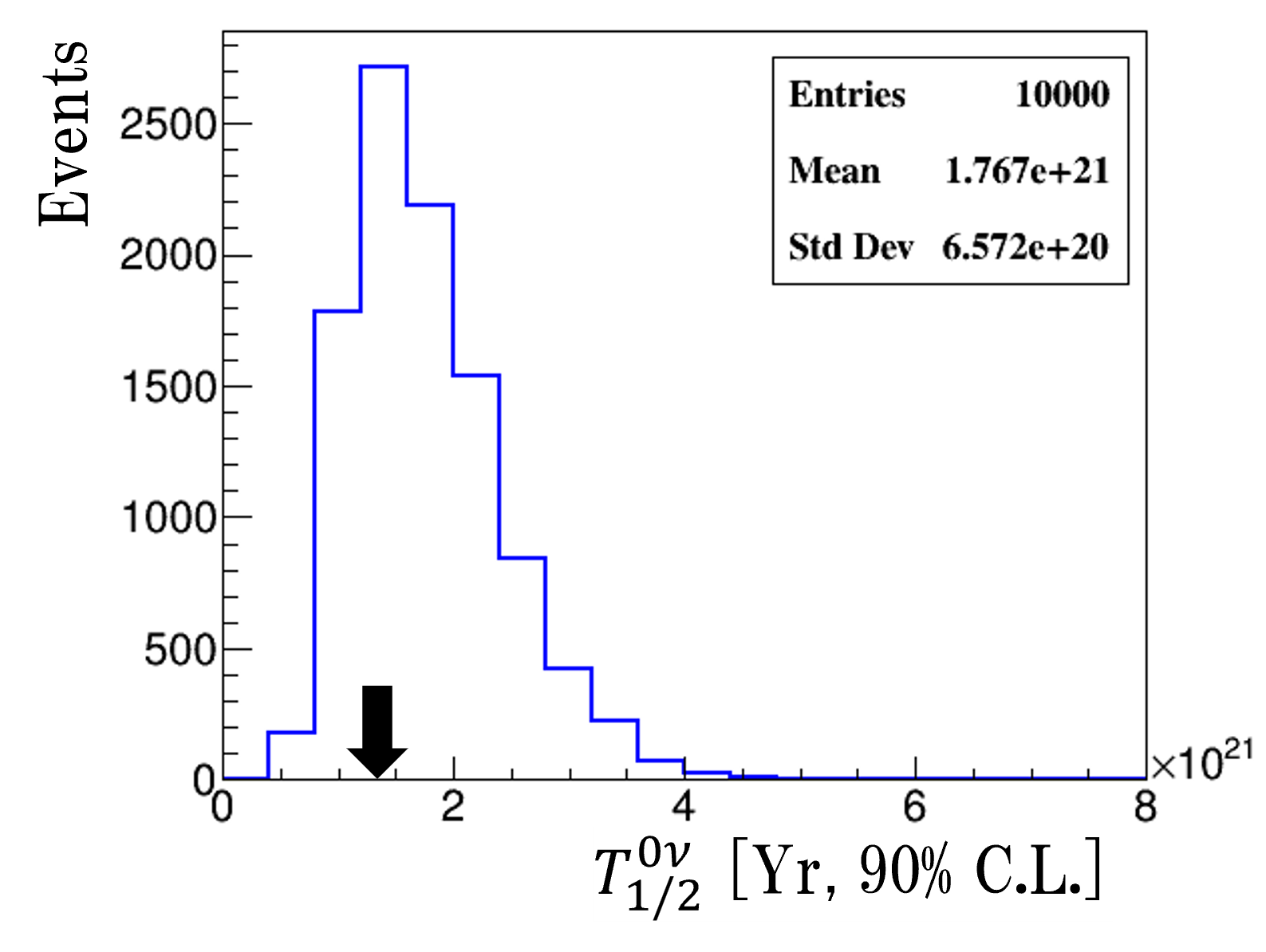}
        \caption{For 0$\nu$2$\beta$. The median is 1.66 $\times$ 10$^{21}$ years.}
        \label{chi2 0v limit}
    \end{subfigure}
    \caption{The distributions of the lower limits of half-lives at 90$\%$ C.L. estimated by Method~2. The black arrows indicate the limits from the previous study \cite{Ukraine}.}
    \label{chi2 limit}
\end{figure}

\noindent The red lines in Fig. \ref{2v chi2} (\ref{0v chi2}) is drawn at $\chi_{2\nu(0\nu)}^{2}$ where Probability equals 0.9, indicating the upper limit of 2$\nu$2$\beta$ (0$\nu$2$\beta$) rate at 90$\%$ C.L.\\
Substituting the 2$\nu$2$\beta$ (0$\nu$2$\beta$) rates evaluated for all the pseudo spectra into $N^{2\nu(0\nu)}_{\rm{obs}}$ in Eq. (\ref{sens}), we obtained the distribution of the lower limits of $T^{2\nu}_{1/2}$ and $T^{0\nu}_{1/2}$ at 90$\%$ C.L., as shown in Figs.~\ref{chi2 2v limit} and \ref{chi2 0v limit}, respectively. The arrows described in Fig. \ref{chi2 limit} represent each previous limit, and the lower limits of $T^{2\nu}_{1/2}$ and $T^{0\nu}_{1/2}$ exceeded these limits for 8,110 and 7,344 sets of pseudo spectra, respectively. In conclusion, we estimated representative (median) values of 2.64 $\times$ 10$^{19}$ years for $T^{2\nu}_{1/2}$ and 1.66 $\times$ 10$^{21}$ years for $T^{0\nu}_{1/2}$ at 90$\%$ C.L., respectively.

Table \ref{Summary} summarizes the lower limits of  $T^{2\nu}_{1/2}$ and $T^{0\nu}_{1/2}$ in Phase~1, estimated by two different approaches, along with the limits from the previous study. As shown in Table \ref{Summary}, both $T^{2\nu}_{1/2}$ and $T^{0\nu}_{1/2}$ were consistently evaluated for Method~1 and Method~2, and they exceed the limits from the previous study.

\begin{table}[!h]
\caption{The lower limits of half-lives estimated for Phase~1 at 90$\%$ (68$\%$) C.L.}
\label{Summary}
\centering
\renewcommand{\arraystretch}{1.5}
\begin{tabular}{llcc}\hline
Experiment & Method & $T^{2\nu}_{1/2}$ [Yr] & $T^{0\nu}_{1/2}$ [Yr] \\ [2ex]\hline
\multirow{2}{*}{Phase 1} & Method 1 & 2.63 (4.32) $\times$ 10$^{19}$ & 1.65 (2.71) $\times$ 10$^{21}$ \\
                        & Method 2 & 2.64 (4.38) $\times$ 10$^{19}$ & 1.66 (2.75) $\times$ 10$^{21}$  \\\hline
Ref. \cite{Ukraine}     & & 1.9 (3.1) $\times$ 10$^{19}$ & 1.3 (2.3) $\times$ 10$^{21}$  \\\hline
\end{tabular}
\end{table}

\section{Discussion}

\subsection{The purity of GAGG crystal required for Phase~2}
\begin{figure}[!h]
    \centering
    \includegraphics[keepaspectratio,scale=0.7]{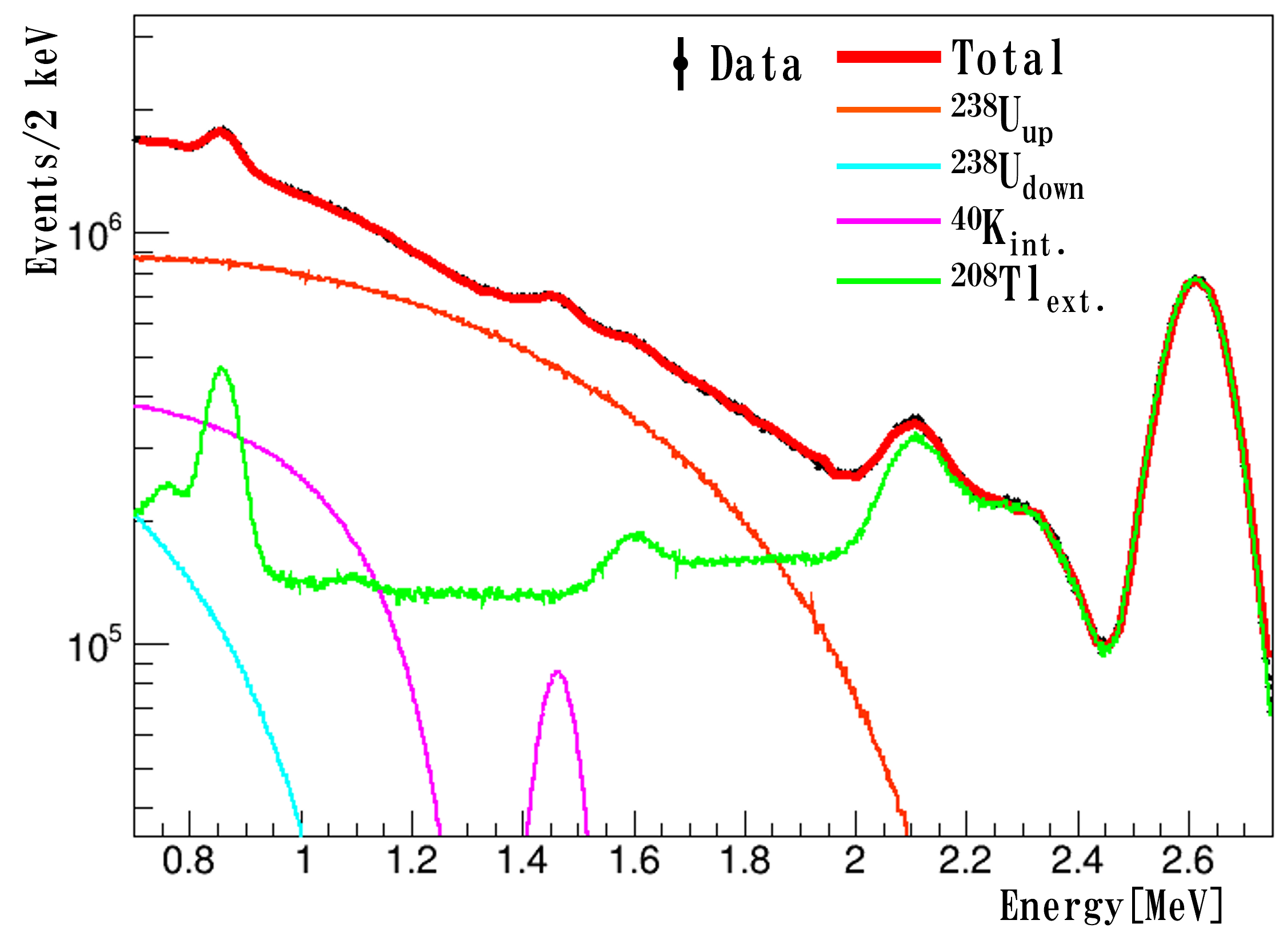}
    \caption{A particular $\beta$($\gamma$)-ray pseudo spectrum for Phase~2 represented by black points, along with the best-fit background model spectra, represented by various color lines. The pseudo spectra were generated by Poisson distribution as described in Section \ref{Sensitivity study}. However, the base model was adjusted to reduce the impurities in GAGG of $^{238}$U$_{\rm{up}}$, $^{232}$Th, and $^{40}$K$_{\rm{int.}}$ by an order of magnitude, and to remove $^{40}$K$_{\rm{ext.}}$ entirely.}
    \label{beta Phase2}
\end{figure}

The goal of Phase~2 is to either discover 2$\nu$2$\beta$ or constrain the theoretical model for NME, by searching for 2$\beta$ with a sensitivity more than 7.4 $\times$ 10$^{20}$ years which is the theoretical prediction \cite{Hinohara} as described in Section \ref{Introduction}. For the 2$\nu$2$\beta$ search, the main radioactive background sources in GAGG are currently 122 $\pm$ 0 mBq/kg of $^{234m}$Pa in $^{238}$U$_{\rm{up}}$, 2.27 $\pm$ 0.02 mBq/kg of $^{228}$Ac in $^{232}$Th, and 39.7 $\pm$ 0.5 mBq/kg of $^{40}$K$_{\rm{int.}}$ (see Fig. \ref{beta bestfit}). In contrast, the Gd$_{2}$O$_{3}$, which is the main raw material for GAGG, has been measured to contain $<$ 16.3 mBq/kg, $<$ 0.96 mBq/kg, and $<$ 2.7 mBq/kg of them, respectively, as measured by a high purity germanium detector \cite{PIKACHU}. Other raw materials have also been confirmed to exhibit significantly higher purity than the crystal itself. These discrepancies suggest contamination from the materials installed in the crystal growing furnace. However, they also indicate the potential to develop GAGG with a purity at least an order of magnitude higher than the current level. If such an appropriate purification can be achieved, $^{40}$K$_{\rm{ext.}}$ would become a dominant background in the 2$\nu$2$\beta$ search. We will address this issue through the strategies described in the next section. \\
Suppose Phase~2 experiment is conducted for 5 years, using 35.7 kg $^{160}$Gd in a hundred of large GAGG crystals with an order of magnitude lower concentrations of $^{238}$U$_{\rm{up}}$, $^{232}$Th, and $^{40}$K$_{\rm{int.}}$ compared to those of high-purity GAGG. Furthermore, assuming we utilize ideal photodetectors without $^{40}$K$_{\rm{ext.}}$ contamination. Under these assumptions, the background spectrum of Phase~2 is expected to be as shown in Fig. \ref{beta Phase2}, which represents one of the $\beta$($\gamma$)-ray pseudo spectrum generated in the same way as described in Section \ref{Sensitivity study}. By the Method~2 analysis for the 10,000 data-set of the Phase~2 pseudo spectra, we obtained 1.17 (1.96) $\times$ 10$^{21}$ years as the lower limit of $T^{2\nu}_{1/2}$ at 90$\%$ (68$\%$) C.L., providing sufficient sensitivity to meet the objectives of Phase~2. 

\subsection{Strategies to reduce the $\gamma$-ray background from $^{40}\rm{K}$ in photodetectors}
\begin{figure}[!h]
 \centering
    \includegraphics[keepaspectratio,scale=0.4]{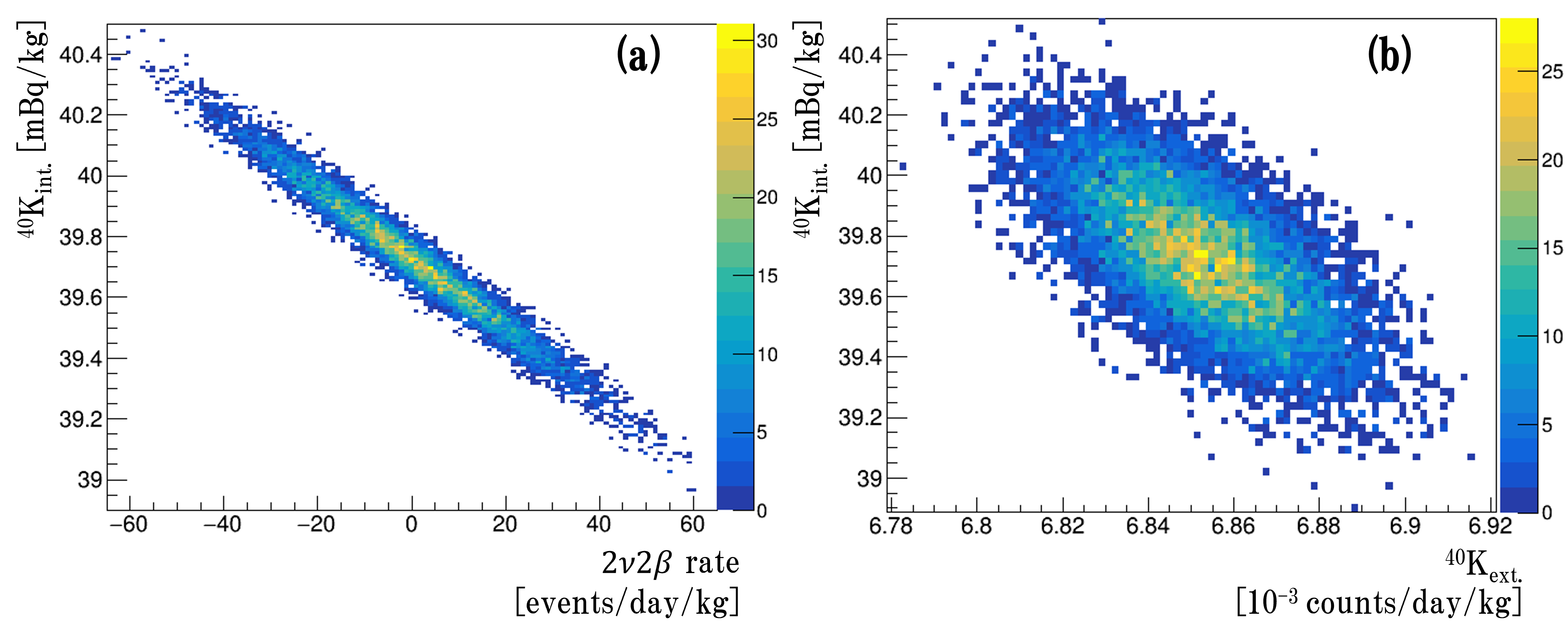}
    \caption{The negative correlations between 2$\nu$2$\beta$ rate and $^{40}$K$_{\rm{int.}}$ (a), and between $^{40}$K$_{\rm{ext.}}$ rate and $^{40}$K$_{\rm{int.}}$ rate (b). These are obtained by fitting 10,000-set of the pseudo spectrum expected in Phase~1 with the background model spectra combined with 2$\nu$2$\beta$ model spectrum.}
    \label{K40}
\end{figure}

Crystal purification, including a reassessment of the crystal growth environment and optimization of growth conditions, will remain as a long-term challenge. However, the reduction of $^{40}$K$_{\rm{ext.}}$ can effectively contribute to the indirect improvement of the 2$\nu$2$\beta$ sensitivity, and the strategies for achieving this are relatively straightforward. \\
$^{40}$K$_{\rm{int.}}$ will be a critical background source in the 2$\nu$2$\beta$ search because the beta-minus decay events of $^{40}$K$_{\rm{int.}}$ with a $Q_{\beta^{-}}$ of 1.31~MeV are distributed within the 2$\nu$2$\beta$ energy region, resembling the shape of the 2$\nu$2$\beta$ spectrum. Indeed, their negative correlation can be seen in Fig. \ref{K40} (a), which shows the 2$\nu$2$\beta$ rate and the $^{40}$K$_{\rm{int.}}$ radioactivity obtained through the best-fit of all the pseudo data. Therefore, an accurate evaluation of $^{40}$K$_{\rm{int.}}$ is desirable for the precise 2$\nu$2$\beta$ analysis. Additionally, $^{40}$K$_{\rm{ext.}}$ is also negatively correlated with $^{40}$K$_{\rm{int.}}$ due to the similarity of the shape of their model spectra, as shown in Fig. \ref{K40} (b), which contributes to the uncertainty of $^{40}$K$_{\rm{int.}}$. In other words, if the detection rate of $^{40}$K$_{\rm{ext.}}$ can be sufficiently suppressed relative to the radioactivity of $^{40}$K$_{\rm{int.}}$, the $^{40}$K$_{\rm{int.}}$ content can be estimated accurately from the photoelectron peak at 1.46~MeV.

Two main strategies are being considered to reduce $^{40}$K$_{\rm{ext.}}$, the $\gamma$-ray background from $^{40}$K in the photodetectors. One is to use low radioactivity PMTs, while the other is to use multi-pixel photon counters (MPPC) to detect the GAGG scintillation. \\
Low radioactivity PMTs, such as the R8778 used in the XMASS-I experiment \cite{r8778}, have been developed for the dark matter search in low background environments. The R6231-100 was measured to contain 5.58 $\pm$ 0.05 Bq/PMT of $^{40}$K, as determined by a high purity germanium detector, while the R8778 contains 140 $\pm$ 20 mBq/PMT of $^{40}$K \cite{r8778}, which can suppress the rate of $^{40}$K$_{\rm{ext.}}$ by more than an order of magnitude compared to the current level. However, the quantum efficiency (Q.E.) of R8778 is optimized for about 178 nm wavelength of liquid xenon scintillation \cite{Xe}. For Q.E. at around 520 nm corresponding to the wavelength of GAGG scintillation \cite{GAGG}, the Hamamatsu R6231-04 can be a suitable candidate because its Bialkali photocathode is sensitive to wavelengths around 300 $\sim$ 650 nm. Furthermore, due to its input window being made of potassium-free glass, the $^{40}$K content is suppressed to 1.15 $\pm$ 0.07 Bq/PMT, as confirmed by measurement using the high purity germanium detector. \\
MPPC can also be considered as a useful photodetector candidate in our experiment because its peak sensitivity wavelength of it is typically around 450 nm. In fact, GAGG scintillation has been successfully read out by an MPPC array \cite{MPPC}. The MPPC is composed of fewer materials than the PMT, and in particular, it does not include glass, which typically contains the majority of $^{40}$K among the PMT components. In this strategy, the MPPC must be arrayed to extend the photosensitive area, thereby covering the cross-sectional area of the GAGG crystal. In addition, the signal readout circuit must be designed to accommodate the arrayed MPPC.

\section*{Acknowledgments}
This work was supported by JSPS KAKENHI Grant No. 22H04570 and 23H01196, and the Inter-University Cooperative Research Program of the Institute for Materials Research, Tohoku University (Proposal No. 202112-RDKGE-0005 and 202012-RDKGE-0016). This experiment was conducted in a section of the area provided by Research Center for Neutrino Science, Tohoku University.





\end{document}